\documentclass[twocolumn,english,prd,superscriptaddress,aps,showpacs,showkeys,amsmath,amssymb,floatfix,nofootinbib]{revtex4-1}
\pdfoutput=1
\usepackage[pdftex]{graphicx}
\usepackage[pdftex]{color}
\usepackage[colorlinks,bookmarks]{hyperref}
\definecolor{linkblue}{rgb}{0,0,0.8}
\definecolor{linkgreen}{rgb}{0,0.5,0}
\definecolor{darkgreen}{rgb}{0,0.4,0}
\definecolor{purple}{rgb}{0.7,0.0,0.4}
\hypersetup{linkcolor=linkblue, citecolor=purple, urlcolor=linkblue}

\usepackage{nicefrac}

\usepackage[T1]{fontenc}
\usepackage[utf8]{inputenc}
\usepackage{lmodern}
\usepackage{babel}
\usepackage{bm}
\usepackage{verbatim}
\usepackage[us]{datetime}
\usepackage{enumerate}

\newcommand{\omegak}{\Omega_{k0}}

\newcommand{\tgl}{\texttt{turboGL}\ }

\begin{document}

\title{Accurate weak lensing of standard candles. I. Flexible cosmological fits}

\author{Valerio Marra}
\affiliation{Institut für Theoretische Physik, Universität Heidelberg, Philosophenweg
16, 69120 Heidelberg, Germany}

\author{Miguel Quartin}
\affiliation{Instituto de Física, Universidade Federal do Rio de Janeiro, CEP
21941-972, Rio de Janeiro, RJ, Brazil}

\author{Luca Amendola}
\affiliation{Institut für Theoretische Physik, Universität Heidelberg, Philosophenweg
16, 69120 Heidelberg, Germany}

\begin{abstract}
With the availability of thousands of type Ia supernovae in the near future the magnitude scatter induced by lensing will become a major issue as it affects parameter estimation. Current N-body simulations are too time consuming to be integrated in the likelihood analyses used for estimating the cosmological parameters. In this paper we show that in the weak lensing regime a statistical numerical approximation produces accurate results orders of magnitude faster. We write down simple fits to the second, third and fourth central moments of the lensing magnification probability distribution as a function of redshift, of the power spectrum normalization and of the present-day matter density. We also improve upon existing models of lensing variance and show that a shifted  lognormal distribution fits well the numerical one. These fits can be easily  employed in cosmological likelihood analyses. Moreover, our theoretical predictions make it possible to invert the problem and begin using supernovae lensing to constrain the cosmological parameters.
\end{abstract}




\maketitle

\section{Introduction}\label{intro}

More than a decade after the discovery of acceleration~\cite{Perlmutter:1998np,Riess:1998cb}, type Ia supernovae (SNe) are still the most accurate tool to map the cosmic expansion at low redshift. Intense search programs for the next years, like DES~\cite{Bernstein:2011zf} and the LSST~\cite{Abell:2009aa} will make available to the cosmologists thousands of new supernovae, enhancing the use of these standard candles as cosmology
probes. With the honor of being the most useful cosmology probe comes
the duty of understanding and reducing the impact of possible systematic
biases~\cite{Amendola:2012wc}, for instance dust extinction, progenitor contamination, intrinsic
variability etc.

Among these systematic effect, perhaps the most interesting from a cosmological point of view is the magnification lensing. The light ray bundle from any source at large redshift in an inhomogeneous universe is of course subject to magnification or demagnification that mixes with the intrinsic source magnitude and change its statistical properties. The intrinsic scatter among supernovae at a given redshift is therefore convolved with the scatter induced by the matter fluctuations along the line of sight. This effect was recently observed with current supernovae~\cite{Kronborg:2010uj,Jonsson:2010wx,Smith:2013bha}, albeit only at around 2$\sigma$ significance. Since the fluctuations themselves depend on the cosmological parameters, the lensing noise is actually an additional signal that can help with parameter estimation. We pursue this avenue in~\cite{Quartin:2013moa}. In the present paper instead we focus on estimating the lensing distortions through the second, third and fourth central moments of the lensing PDF.

The lensing probability distribution function (PDF) is the fundamental quantity describing the statistical (de)magnification of distant sources by large-scale
inhomogeneities.
After some pioneering work (e.g. Ref.~\cite{Frieman:1996xk}),
various approaches have been followed so as to compute the lensing PDF relative
to a given cosmology.
A first approach (e.g.~\cite{Munshi:1999qw, Valageas:1999ir, Wang:2002qc,
Das:2005yb}) relates a ``universal'' form of the lensing PDF to the variance of
the convergence, which is fixed by the amplitude $\sigma_{8}$ of the power
spectrum. Moreover the coefficients of the proposed PDF may be trained on a grid
of N-body simulations. A second approach (e.g.~\cite{Holz:1997ic,
Bergstrom:1999xh, Holz:2004xx, Hilbert:2007ny, Hilbert:2007jd, Takahashi:2011qd,
Bolejko:2012ue}) is to build a model of the universe, e.g.~by means of an N-body
simulation, and directly compute the relative lensing PDF, usually through
time-consuming ray-tracing techniques. The flexibility of this method is
therefore penalized by the increased computational time.

In order to combine the flexibility in modeling with a fast performance in
obtaining the lensing PDF the stochastic gravitational lensing (sGL) method has
been introduced in
Refs~\cite{Kainulainen:2009dw,Kainulainen:2010at,Kainulainen:2011zx}. The sGL
method is based on the weak lensing approximation and generating stochastic
configurations of inhomogeneities along the line of sight. For instance, its
speed allows one to include the full (cosmology-dependent) weak-lensing effects
in the supernova analysis.  This was first carried out in~\cite{Amendola:2010ub}
for a simple toy-model, consisting of a universe populated by a distribution of
large halos, all with the same mass, ${\mathcal{O}}(10^{14}M_\odot)$.

The outline of this paper is as follows. In Section \ref{setup} we will describe the universe model we adopt and briefly review the very basics of sGL. In Section \ref{moments} we will give our results as far as the lensing moments are concerned, while in Section \ref{sec:logn} we will show how to reconstruct the lensing distribution from the latter moments. Finally, we will conclude in Section \ref{sec:conclusions}.
In the Appendixes \ref{sec:app-review} and \ref{sec:errors}, we discuss some properties of lensing and the errors intrinsic in our approach, respectively.

\section{Setup}\label{setup}

\subsection{Universe model}

We consider homogeneous and isotropic Friedmann-Lemaître-Robertson-Walker (FLRW)
background solutions to Einstein's equations, on top of which we add matter
perturbations describing virialized halos. We assume as fiducial a model
the non-flat $w$CDM, for which the parameters
describing the background are the present-day expansion rate $h$, matter density
parameter $\Omega_{m0}$, curvature parameter $\omegak$ and
dark-energy equation of state parameter~$w$.

We will model the matter contrast $\delta_{M}(r,t)$ according to the so-called
``halo model'' (HM) (see, for example, \cite{Neyman:1952, Peebles:1974,
Scherrer:1991kk, Seljak:2000gq, Ma:2000ik, Peacock:2000qk, Scoccimarro:2000gm,
Cooray:2002dia}), where the inhomogeneous universe is approximated as a
collection of different types of halos whose positions satisfy the linear power
spectrum.
More precisely, the halo model assumes that on small scales (large wave numbers
$k$) the statistics of matter correlations are dominated by the internal halo
density profiles, while on large scales the halos are assumed to cluster
according to linear theory. In other words, the nonlinear evolution is assumed
to produce only concentrated halos. The model does not include intermediate
extended structures such as filaments and walls. See \cite{Kainulainen:2010at}
for an improved halo model that also includes filaments.

The parameters describing the spectrum of perturbations are, besides the ones relative to the background, the power spectrum
normalization $\sigma_{8}$, spectral index $n_{s}$ and baryon density parameter
$\Omega_{b 0}$. For the transfer function $T(k)$ we use the fit provided by the
Equations (28)--(31) of Ref.~\cite{Eisenstein:1997ik}.
See~\cite{Kainulainen:2010at} for more details about the power spectrum used.

Regarding the modeling of halos, we use the halo mass function given in Eq.~(B3) of Ref.~\cite{Jenkins:2000bv}, which is defined relative to a
spherical-overdensity halo finder and has a good degree of universality~\cite{White:2002at}.
Our results are not very sensitive with respect to the mass function adopted; if for example the halo mass function of Ref.~\cite{Courtin:2010gx} is used, then approximately 5\% higher lensing moments will be obtained.
The halo profiles are modeled according to the Navarro-Frenk-White (NFW) profile \cite{Navarro:1995iw}, which is able to model both galaxy-sized halos and superclusters with a appropriately chosen concentration parameter  $c(M,z)$. The concentration parameter depends on the cosmology and we use the universal model proposed in Ref.~\cite{Zhao:2008wd}, which allows for an accurate determination of a halo's concentration by relating the latter to the time at which the halo's
main progenitor first assembled 4\% of its final mass. We however neglect correlations in the halo positions. As shown in~\cite{Kainulainen:2010at,Kainulainen:2011zx}, this should be a good
approximation for the redshift range of $z \lesssim 1$ in which we are mainly interested in this paper.

Summarizing, our universe model, which is determined by the vector of parameters
$\{ h, \Omega_{m0}, \omegak, w, \sigma_{8}, n_{s}, \Omega_{b0} \}$,
approximates in a consistent way the real universe at the nonlinear scales of
clusters, the ones relevant for lensing of standard candles.

\subsection{Lensing model} \label{sec:model}

As we will now see, the sGL method predicts the statistical distribution of the
lens convergence $\kappa$, which in the weak-lensing approximation is given by
the following integral evaluated along the unperturbed light
path~\cite{Bartelmann:1999yn}:
\begin{equation} \label{eq:kappa}
    \kappa(z_{s})=\int_{0}^{r_{s}}dr \, \rho_{MC} \, G(r,r_{s})\,\delta_{M}(r,t(r))
\end{equation}
where the quantity $\delta_{M}(r,t)$ is the local matter density contrast, the
density $\rho_{MC} \equiv a_0^3 \, \rho_{M0}$ is the constant matter density in
a co-moving volume, and we defined the auxiliary function
\begin{equation} \label{opw}
G(r,r_{s})=  \frac{4\pi G}{c^2 \, a}  \; \frac{f_{k}(r)f_{k}(r_{s}-r)}{f_{k}(r_{s})} \,,
\end{equation}
which gives the optical weight of a matter structure at the comoving radius $r$.
The functions $a(t)$ and $t(r)$ are the scale factor and geodesic time for the
background FLRW model; $r_{s}=r(z_{s})$ is the comoving position of the source
at redshift $z_{s}$; and finally \linebreak[4]
$f_{k}(r)=\sin(r\sqrt{k})/\sqrt{k},\; r,\;\sinh(r\sqrt{-k})/\sqrt{-k}\,$
depending on whether the curvature k is $>,=,<0$, respectively.
At linear level, 
the shift in the distance modulus caused by lensing is expressed in terms of the
convergence only:
\begin{equation} \label{eq:dm}
\Delta m(z) \simeq  5 \log_{10}\big [1-\kappa(z) \big] \simeq - \frac{5}{\log 10} \; \kappa(z) \,.
\end{equation}
Eqs. (\ref{eq:kappa}) and (\ref{eq:dm}) illustrate how for a lower-than-FLRW column density the light is demagnified, while in the opposite case it is magnified.
Also note that Eq.~(\ref{eq:kappa}) is only valid for point sources (smoothing angle $\theta =0$).
Our approach is directly based on convergences; however, we will present our results in magnitudes as this is the standard choice, that is we will multiply  convergences by the factor $- 5/\log 10 $.

Eq.~(\ref{eq:kappa}) connects the statistical distribution of matter to the
statistical distribution of convergences.
The sGL method for computing the lens convergence is based on generating random
configurations of halos along the line of sight and computing the associated
integral in Eq.~(\ref{eq:kappa}) by binning into a number of independent lens
planes. Because the halos are randomly placed their occupation numbers in
parameter-space volume cells follow the Poisson statistics. This allows
rewriting Eq.~(\ref{eq:kappa}) as a sum over these cells characterized by the
corresponding Poisson occupation numbers.
The various individual contributions to the convergence are then additively
combined. By generating many halo configurations one can sample the convergence
PDF.
A detailed explanation of the sGL method, which can also model filamentary
structures confining halos, can be found
in~\cite{Kainulainen:2011zx,Kainulainen:2010at,Kainulainen:2009dw} and a
publicly-available numerical implementation, the \tgl package, at
\href{http://www.turbogl.org/}{turbogl.org}. We note that
although \tgl is not strictly speaking a ray-tracing procedure it emulates it by
constructing many halo realizations along a given direction. To obtain the
results in this paper we ran \tgl with a statistics of $1.5 \cdot 10^{6}$
``light rays''. In other words, we shoot 1.5 million light rays for every point in the cosmological parameter space.

\begin{figure*}
\begin{center}
\includegraphics[height=5.4 cm]{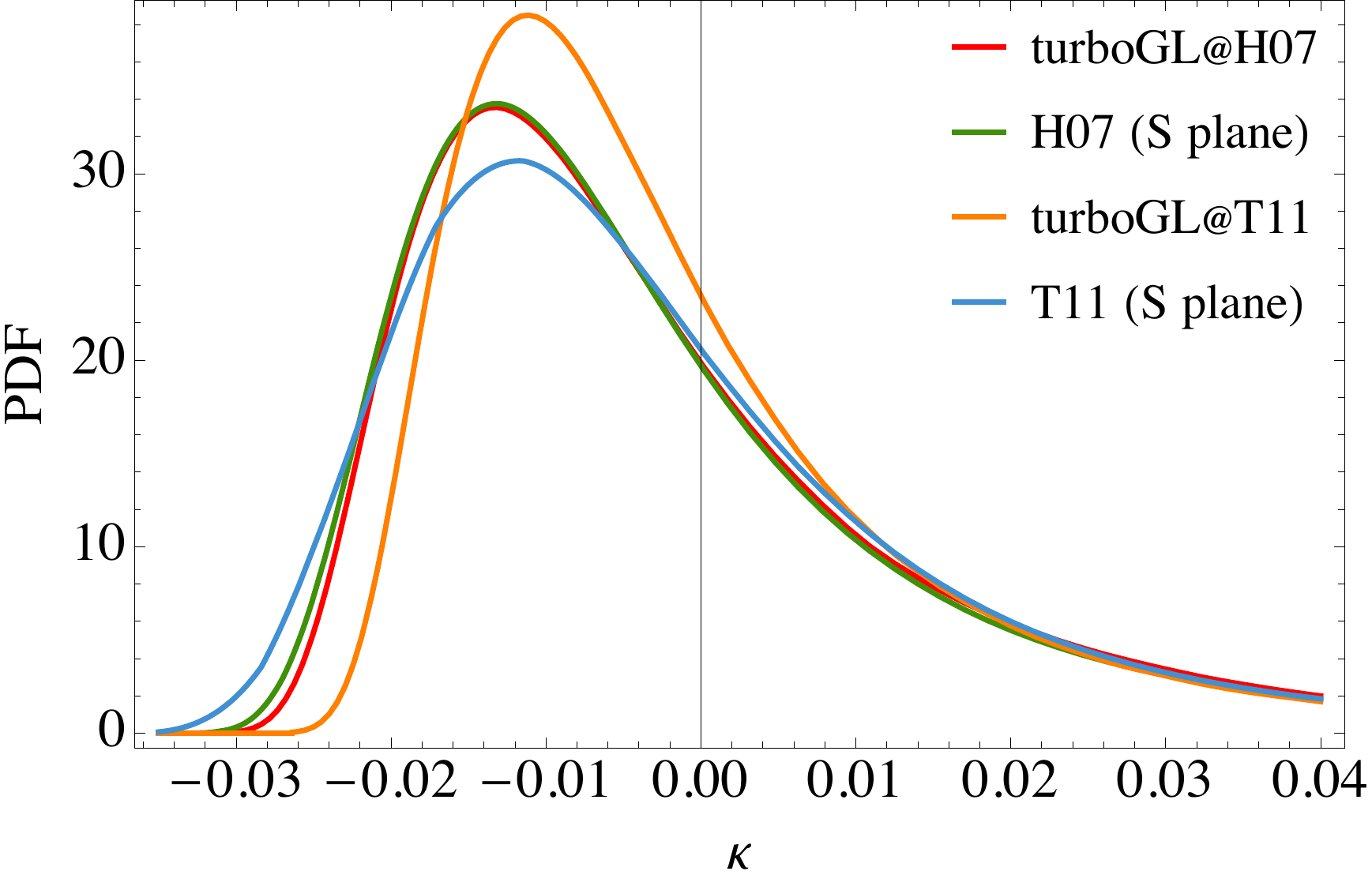}
\qquad
\includegraphics[height=5.4 cm]{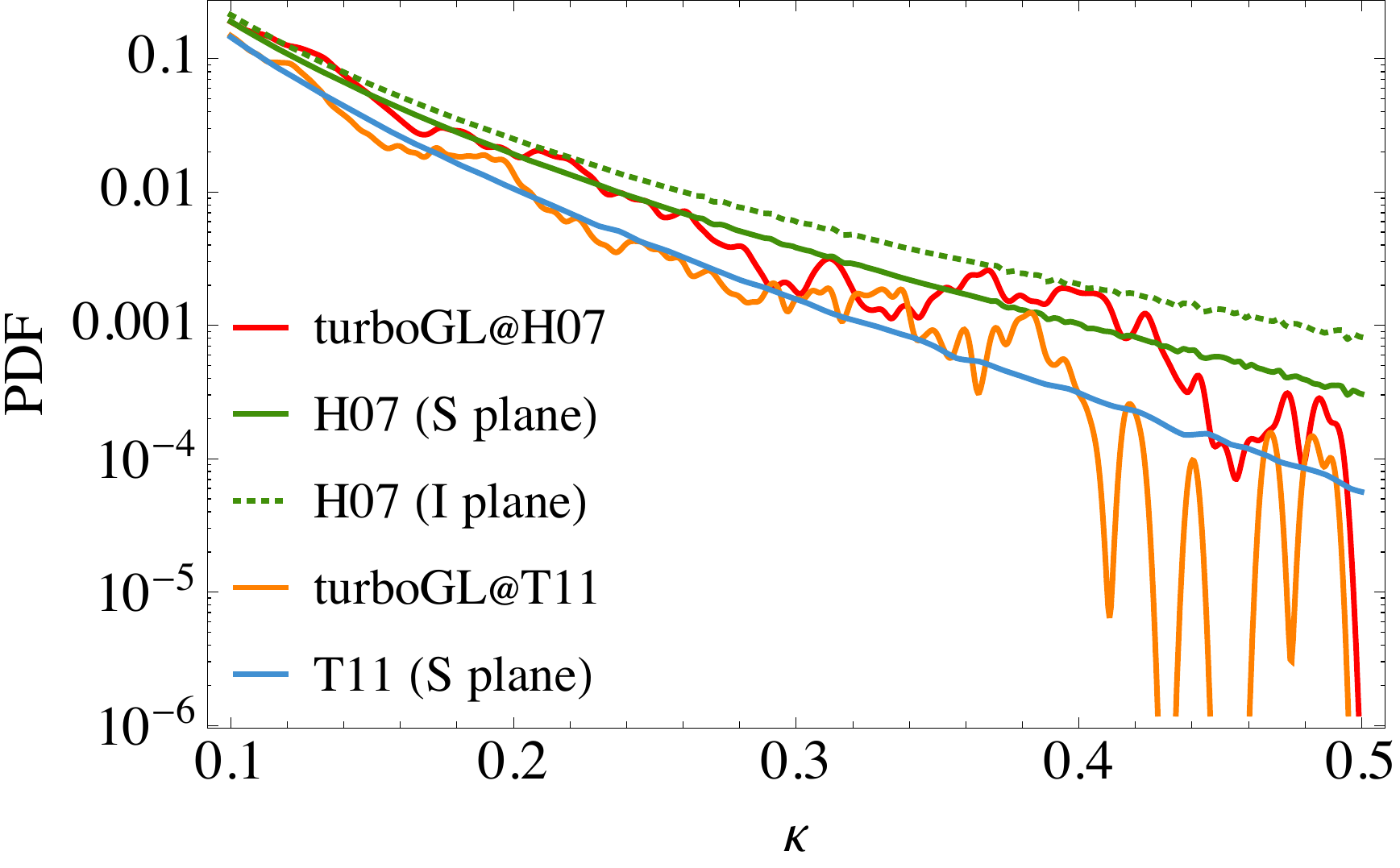}
\caption{Comparison of the convergence PDF's from \tgl\cite{Kainulainen:2009dw,Kainulainen:2010at}
and ray-tracing from two N-body simulations, dubbed H07~\cite{Hilbert:2007ny,Hilbert:2007jd}
and T11~\cite{Takahashi:2011qd}. The former is based
on the Millennium Simulation (at $z=1.08$), the latter on a more
recent simulation (at $z=1$). \emph{Left}: slight discrepancy at
negative $\kappa$; although the cosmological parameters are different
on both simulations, the discrepancy is relevant. \emph{Right}: zoom
on the high-magnification tail and comparison of image and source
plane PDF's. See Section \ref{compase} and Appendix~\ref{sec:app-review} for more details.}
\label{kPDF}
\end{center}
\end{figure*}

Overall, we estimate that our results can be relied upon at the level of $\sim$10\%. We discuss in more detail in Appendix~\ref{sec:errors} the full error budget regarding this estimate and the approximations of this setup.

\section{Moments of the lensing PDF}
\label{moments}

The modeling described in the previous Section and implemented in (the latest)
version 2.0 of \tgl\!\!\!, although many orders of magnitude faster than running
an N-body simulation and relative ray-tracing routine, still needs about 15--20
seconds to generate a PDF with a statistics of $10^{5}$ using one core of a CPU
at 2--3 GHz.
Therefore, it can be computationally expensive to run sGL on a grid or
Monte-Carlo for many points. It is therefore convenient
to create shortcuts which could allow for a fast estimation of the weak
lensing distortions in different cosmologies. In fact, as discussed before,
different shortcuts
to the full problem were proposed in the literature in recent years, see
e.g.~\cite{Munshi:1999qw, Valageas:1999ir, Wang:2002qc, Das:2005yb}. These,
however, usually take the form of a simplified algorithm to compute the lensing
PDF, the application of which is not completely straightforward and still adds
some computational complexity to any likelihood estimation.

Here we take a different approach. Instead of giving a prescription so as to
obtain the lensing PDF, we focus on the information condensed into its central
moments  $\mu_{2-4}$, which are defined as
\begin{equation}
    \mu_n\equiv\left\langle \left(X - \langle X\rangle\right)^n \right\rangle
\,,
\end{equation}
and are directly related to the variance, skewness and kurtosis of the PDF. We
provide below analytical fitting functions for these three moments with respect
to the most relevant cosmological parameters, to wit $\sigma_{8}$ and
$\Omega_{m0}$, as well as redshift, all for the very broad ranges listed in (\ref{parange}).

There are a number of advantages in this approach. First, as will be shown, it
makes it very straightforward to include lensing in any likelihood analysis. Second,
it adds almost no computational complexity. Third, the moments may be directly
related to observables. For instance, if the supernovae are intrinsically
gaussian, the skewness of the lensing PDF would be exactly the skewness of the
final convolved distribution. In other words, our estimation of the skewness at
a given redshift would be the skewness one would measure in the distributions of
standard candles in a thin bin at that redshift.
This in turn also allows one to quickly get answers to questions regarding what
might be called the inverse problem of standard candle lensing: how can a
measurement of the actual supernova PDF (or of any other standard candle
candidate) in different redshift bins constrain cosmological parameters such as
$\sigma_{8}$ and $\Omega_{m0}$. We analyse this latter prospect in detail in a
forthcoming companion publication~\cite{Quartin:2013moa}, and instead focus here
on the moments themselves and how to use them.

\subsection{Comparison with previous results} \label{compase}

We will now compare the PDF generated by \tgl with previous results obtained by
shooting rays through the matter field of N-body simulations. In particular we
will consider the findings of Refs.~\cite{Hilbert:2007ny, Hilbert:2007jd}
(dubbed H07) which are based on the $\Lambda$CDM model of the Millennium
Simulation~\cite{Springel:2005nw} consistent with the 1-year WMAP
results~\cite{Spergel:2003cb} and the ones of Ref.~\cite{Takahashi:2011qd}
(dubbed T11) which is based on an N-body simulation consistent with the more
recent 5-year WMAP results~\cite{Komatsu:2008hk}.

Figure~\ref{kPDF} shows an overall agreement between the convergence PDF's
obtained with \tgl and through ray tracing. The left panel shows a very good agreement between the minimum magnification of \tgl and H07.
The agreement is somewhat worse for the case of T11, which gives a minimum
magnification lower than both \tgl and H07. Regarding H07, this discrepancy
could be due to the different smoothing adopted of the matter field~\cite{Takahashi:priv}. Regarding \tgl\!\!, it could be due to the use of the halo model for the matter field; indeed, as shown in~\cite{Kainulainen:2010at}, the minimum magnification shifts to more negative values if one includes filamentary structures.
It thus seems that \emph{for very small convergences}, the different limitations of \tgl and H07 produce the same results.

The \tgl code can include filaments in the modeling. In the present paper we are interested in the cosmology dependence of lensing; therefore we cannot include filaments as at present there are not available models able to predict the filament mass function and, most importantly, the filament properties (the equivalent of the NFW profile for halos).
Consequently, their use is only limited to comparison with a given $N$-body simulation from where -- as it was done in~\cite{Kainulainen:2010at} -- the properties of the filaments can be inferred.
We will therefore restrict our results to the case in which inhomogeneities are modeled with halos only, like in the halo model discussed above. This level of modeling, as shown below, gives a very good accuracy for $z<1$.

We should stress at this point that while \tgl is limited to the weak-lensing regime, the analyses of H07 and T11 consider full nonlinear lensing. As discussed in the Appendix~\ref{sec:app-review}, one can compute the distribution of lensing distortions in either the source (S) or image (I) plane. The reference plane used by the sGL method and thus by \tgl is the S plane, and we will focus on that plane since it is the one in which lensing is directly related to observed quantities. Strictly speaking, in the deep weak lensing limit of $\kappa \ll 1$ this distinction is irrelevant. However, as we will consider convergences up to $\kappa_{\rm cut}=0.35$, the output of \tgl should be compared to results relative to the S plane. The difference between the PDF's in the two planes is substantial only in the high-magnification tail, as shown in the right panel of Figure~\ref{kPDF}. There it is shown how \tgl agrees remarkably well with both T11 and H07 up to a $\kappa_{\rm cut}=0.35$. In fact, the noisy oscillations at high convergences can be made better by hust using higher statistics (here we plot results with the statistics employed throughout this work -- see Section~\ref{sec:model}). If one considers the original H07 result (which was computed in the image plane) a small disagreement between \tgl is found. In summary, \emph{for small to medium convergences}, \tgl is in very good agreement with \emph{both} T11 and H07, as long as one considers the source plane.

\begin{figure}
    \includegraphics[width=1.06\columnwidth]{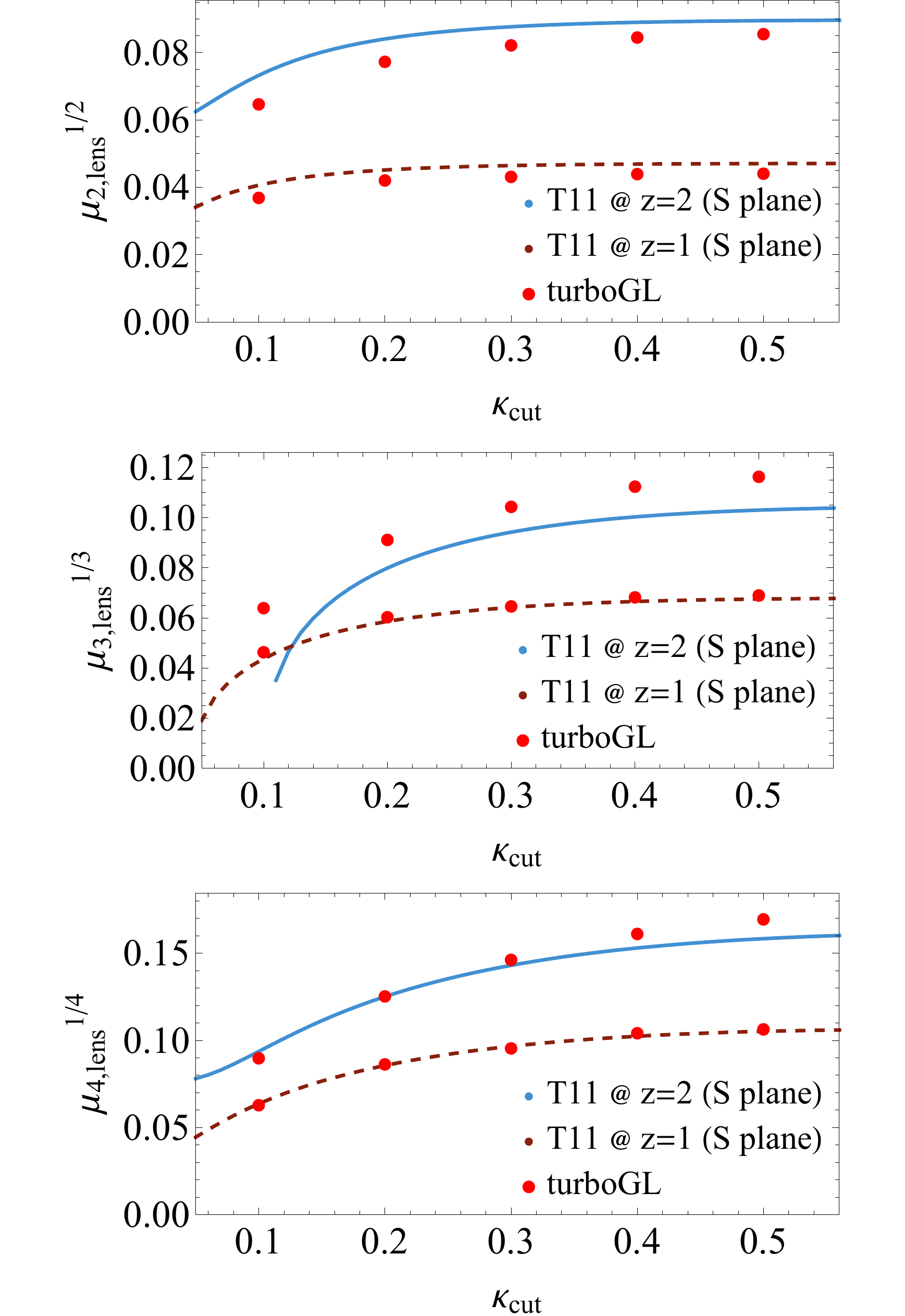}
    \caption{Comparison between the N-body lensing PDF of T11~\cite{Takahashi:2011qd}
    (in the source plane) and \tgl for different choices of $\kappa_{\rm cut}$
    and for both $z=1$ and $z=2$. $\kappa_{\rm cut}$ is the convergence above which we set the PDF to zero. The agreement is very good for $z=1$ but gets slightly worse for the higher moments at higher redshifts, especially for $\mu_{3}$. Since most supernovae lie in the $z<1$ region, we conclude that we can rely on \tgl to very good accuracy. We have chosen $\kappa_{\rm cut}=0.35$ for our fits. All plots are in units of magnitudes, see Eq.~\eqref{eq:dm}. See Section \ref{compase} for more details.}
    \label{fig:takahashi-vs-tGL}
\end{figure}

In this paper we are mainly interested in the moments of the lensing PDF. We
will now test, using T11 as benchmark, the performance of \tgl as far as the
moments are concerned. We will not further consider H07 as we will restrict our
analysis to the lensing PDF in the source plane. Figure~\ref{fig:takahashi-vs-tGL}
shows the comparison of \tgl and T11 for the second-to-fourth central moments
$\mu_{2-4, \text{lens}}$ at the two source redshifts of $z=1$ and $z=2$.\footnote{To ease the comparison we will plot the absolute module of the third moment.} The
plots are with respect to $\kappa_{\rm cut}$, the convergence above which we set
the PDF to zero. The plots show a very good agreement for $z=1$. The agreement
gets slightly worse at $z=2$: this could be due to the fact that the halo model
works less well at the higher redshifts at which non-virialized structures
(neglected in the halo model) play a more important role.
However, since most supernovae lie in the $z<1$ region, we can rely on \tgl to
very good accuracy. Based on these results, we will adopt in the following the
value $\kappa_{\rm cut}=0.35$.

This can be taken as a somewhat conservative value, since based on
Figure~\ref{fig:takahashi-vs-tGL} higher values seem to be in very good agreement
for $z \leq 1$. We nevertheless want to provide estimates for higher redshifts
and, more importantly, for different cosmologies, for which N-body ray tracing
lensing PDF's are currently not available. Although we do not expect a
substantially different agreement for different values of $\Omega_{m0}$ or
$\sigma_8$, we currently cannot test this with N-body simulations.

\begin{figure*}
    \begin{center}
    \includegraphics[width=2.05 \columnwidth]{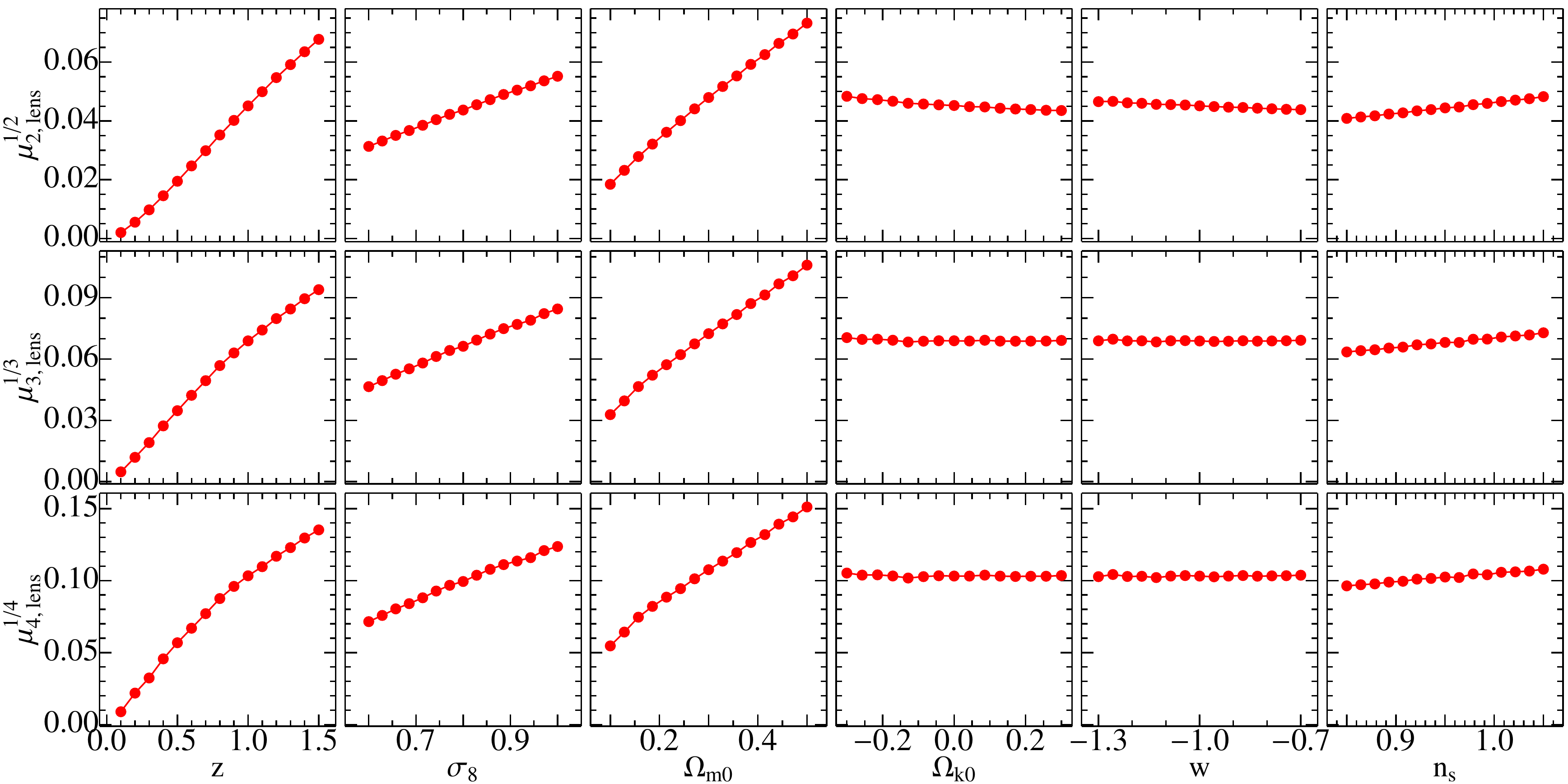}
    \caption{Second-to-fourth central moments  $\mu_{2-4, \text{lens}}$ of the
convergence PDF as a function of redshift and five different cosmological
parameters $\{\sigma_{8},\Omega_{m0},\omegak,w,n_{s}\}$, changing one
at a time while keeping all the others fixed at the 9-year WMAP-only
flat-$\Lambda$CDM best-fit values~\cite{Hinshaw:2012aka} and $z=1$. As can be
seen, apart from $z$ it is clear that $\sigma_{8}$ and $\Omega_{m0}$ are the
most relevant quantities, whilst the dependence on $\omegak$, $w$ and
$n_{s}$ is almost negligible. A value of $\kappa_{\rm cut}=0.35$ has been
adopted, and all plots are in units of magnitudes, see Eq.~\eqref{eq:dm}.
See Section \ref{sec:other-params} for more details.}
    \label{fig:other-params}
    \end{center}
\end{figure*}

Finally, we note that one in principle could try and implement a $\kappa_{\rm
cut}$ on the real data, by excluding the (very rare) standard candles for which
there is independent evidence of either strong or ``medium'' lensing. In fact,
using the full lensing PDF of T11 we estimate that less than 1 in 20000
supernovae at $z = 1$ should have $\kappa_{\rm cut}>0.35$ (and thus
magnification larger than $1.8$), so a more careful check of high magnification
candidates might be feasible (see for instance~\cite{Kronborg:2010uj}).
Incidently, there was a recent claim in the literature~\cite{Quimby:2013lfa} of
a supernova with a huge magnification $\sim$30; the probability for this to
happen due to halos is less than one in a million, and in fact the authors
suggest that the lens might be a compact object such as a black hole. An
exclusion of high magnification events is thus important if one wants to use the
higher moments, as they are the most sensitive to these. We nevertheless do not
investigate further these complications in this work.

\subsection{Dependence on cosmological parameters}
\label{sec:other-params}

In this Section we will quantify how strongly lensing depends on the relevant cosmological parameters. Figure~\ref{fig:other-params} shows the change in the second-to-fourth central moments  $\mu_{2-4, \text{lens}}$ as a function of redshift and five different cosmological parameters $\{\sigma_{8},\Omega_{m0},\omegak,w,n_{s}\}$,
changing one at a time while keeping all the others fixed at the 9-year
WMAP-only fiducial values~\cite{Hinshaw:2012aka} and $z=1$. As we described above, we are assuming a non-flat $w$CDM model throughout this paper. It is clear from Figure~\ref{fig:other-params} that $\sigma_{8}$ and $\Omega_{m0}$ are the dominant parameters, consistently producing variations $\sim$10 larger in observationally comparable ranges. This is what one would expect, as lensing is mostly governed by the amount of matter and related clustering. The dependence on $\omegak$, $w$ and $n_{s}$ (and on $h$ and $\Omega_{b0}$, which were also computed but are not shown in Figure~\ref{fig:other-params}) is almost negligible.
The dependence on $w$ is weak because most of the lensing comes from halos ($M \sim 10^{13} M_{\odot}$) which are not substantially affected by a 30\%-different dark-energy equation of state.
Based on these findings we decided to consider in the following only the dependence of lensing on $z$, $\sigma_{8}$ and $\Omega_{m0}$.



\bigskip

\subsection{Fitting functions for the moments}
\label{sec:fits}

\begin{figure*}[t]
    \begin{centering}
    \includegraphics[width=2\columnwidth]{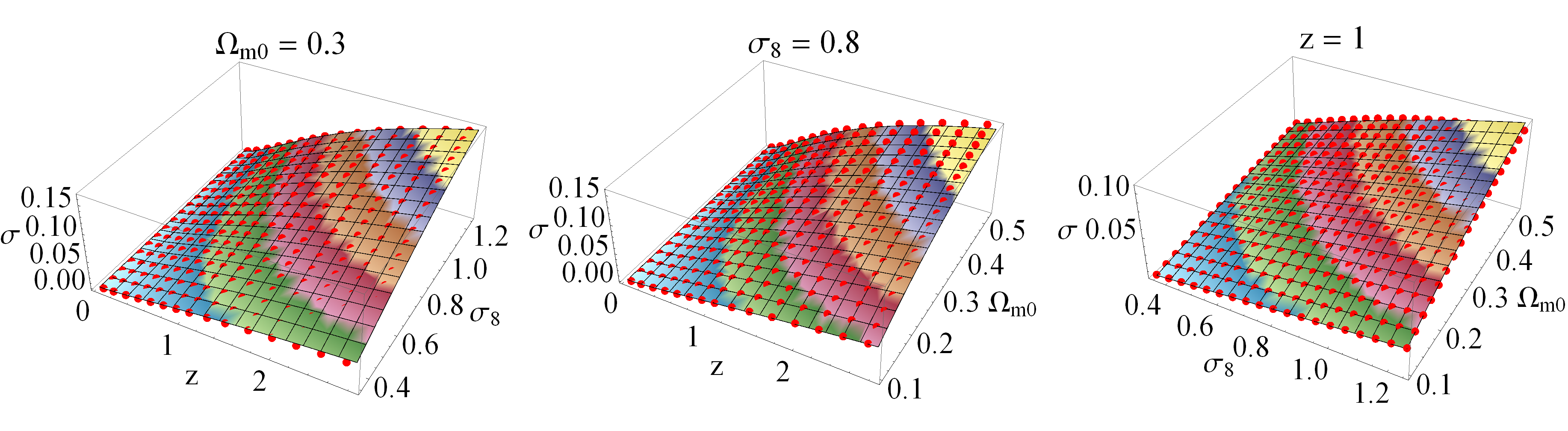}
    \includegraphics[width=2\columnwidth]{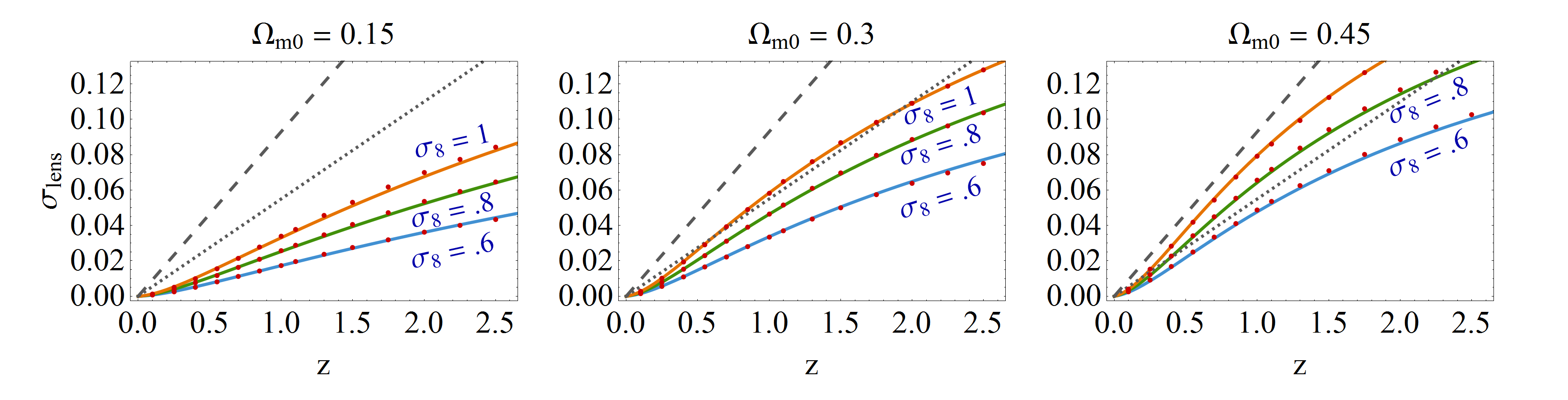}
    \caption{Dependence of the weak-lensing standard deviation (in magnitudes, see Eq.~(\ref{eq:dm})) with respect to $\{z,\,\sigma_{8},\,\Omega_{m0}\}$ from the numerical results (red dots) and the analytical fitting functions (solid surfaces/lines). \emph{Top:} 3D plots covering the full range of validity of the fitting functions. \emph{Bottom:} same but depicted in 2D plots, with the addition of the fits found in~\cite{Holz:2004xx} (gray dashed line) and~\cite{Jonsson:2010wx} (gray dotted line). Note that our fits approximate the numerical results very well in the whole parameter space and are in good agreement with~\cite{Jonsson:2010wx}.}
    \label{fig:fits-sigma} %
    \par\end{centering}
\end{figure*}

\begin{figure*}[t]
    \begin{centering}
    \includegraphics[width=2\columnwidth]{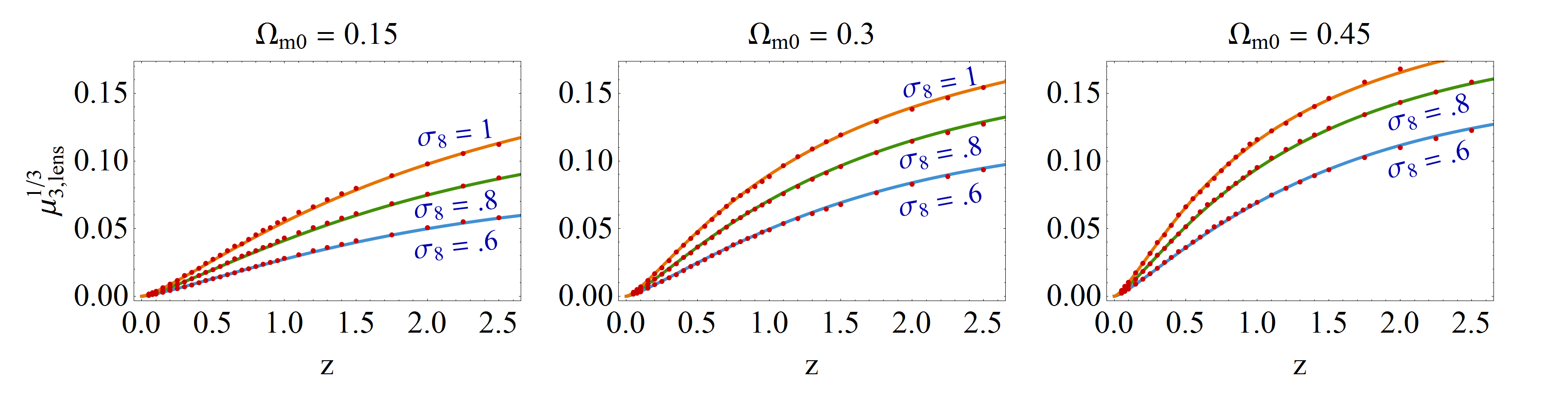}    \includegraphics[width=2\columnwidth]{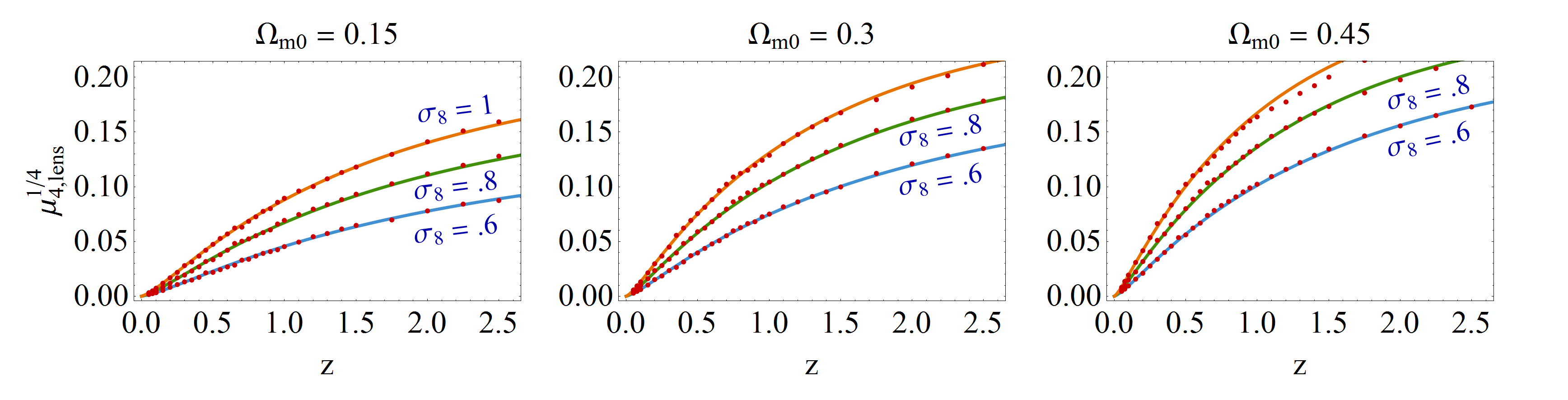}
    \caption{Same as the bottom plot of Figure~\ref{fig:fits-sigma} for the case of the third and fourth moments. Once again, the fitting functions approximate the numerical results very well in the whole parameter space. Note that the use of $\mu_n^{1/n}$ instead of just $\mu_n$ makes it clear that the first moments all have the same order of magnitude.}
    \label{fig:fits-mu3-mu4} %
    \par\end{centering}
\end{figure*}

Here we give analytical fitting functions for the second-to-fourth central
moments $\mu_{2-4, \text{lens}}$ as a function of
$\{z,\,\sigma_{8},\,\Omega_{m0}\}$ which are very accurate in the domain
\begin{equation}
\begin{aligned} \label{parange}
0\,\leq\, & \;\;\;\,z\;\;\,\leq\,3 \,,   \\
0.35\,\leq\, &\;\;\,\sigma_{8}\;\,\leq\,1.25  \,, \\
0.1\,\leq\, & \,\Omega_{m0}\,\leq\,0.52 \,.
\end{aligned}
\end{equation}
In fact, in this entire domain, the average RMS error is only $4\%$ for all
three moments.  Using \emph{magnitudes} (see Eq.~(\ref{eq:dm})), the fitting
formulae are:
\begin{widetext}
\begin{align}
    \sigma_{\rm lens}(z, \sigma_8, \Omega_{m0}) &= \frac{0.0004 - 0.00176 \sigma_8 + (-0.035 + \sigma_8 \,\Omega_{m0} + 0.0453 \sigma_8)z} {\big(2.19  +  \sigma_8^2\big) \Omega_{m0} z +  3.19\,\exp\!\big[0.365/(0.193 + z)\big]}\, ;  \label{2fit}\\
    \mu_{3,{\rm lens}}^{1/3}(z, \sigma_8, \Omega_{m0}) &= \frac{\sigma_8^2\,\Omega_{m0}\,z^2}{\sigma_8 \sqrt{z} + 1.1 z +\big(4.24 \sigma_8^2 - \Omega_{m0}^2\big) \Omega_{m0}\,z^2 +0.118 ( 1-  \sigma_8)z^3 }\,; \label{3fit} \\
    \mu_{4,{\rm lens}}^{1/4}(z, \sigma_8, \Omega_{m0}) &= \frac{(-0.029 + 0.1 \sigma_8 + 0.47 \Omega_{m0} \sigma_8) z}{
    \exp\!\left[(-0.029 + 0.1 \sigma_8 + 0.47 \Omega_{m0} \sigma_8) z + \frac{0.021}{0.018 + \Omega_{m0} \sigma_8 z}\right] + 0.3 z}\,. \label{4fit}
\end{align}
\end{widetext}

Figure~\ref{fig:fits-sigma} depicts the fits for $\sigma_{{\rm lens}}$
for different parameters. The first row cover the full range of validity
of the fitting functions. The middle row represents the same but depicted
in 2D plots, with the addition of the Holz \& Linder fit of $\sigma_{HL}= 0.093 \, z$ \cite{Holz:2004xx} (gray dashed line) and of best fit to the real supernovae data $\sigma_{J10}= 0.055 \, z$ found by Jönsson \emph{et al.}~\cite{Jonsson:2010wx} (gray dotted line). Figure~\ref{fig:fits-mu3-mu4} similarly shows the fits for $\mu_3^{1/3}$ and $\mu_4^{1/4}$. As can be seen, all three central moments are of the same order of magnitude. For all three moments the fit are almost always within 5\% of the numerical values and have an average RMS error of 4\% (see Appendix~\ref{sec:errors} for more details).
These fits improve considerably upon the results of Refs~\cite{Bernardeau:1996un,Hamana:1999rk}.

We would like to point out that the convergence PDF has also a small but nonzero mean.
Although by construction the sGL method produces convergence distributions with zero mean (see Eq.~(\ref{eq:kappa})), the latter can be approximated (in the source plane) by
\begin{equation}
  \langle\kappa\rangle\,\simeq\,-2\langle\kappa^{2}\rangle\,\simeq -2 \sigma_{\rm lens}^2 \,,
\end{equation}
see Appendix~\ref{sec:app-review} for more details.

Finally, the statistics of moments of a given order depends on moments of even higher order. Therefore, the results of Eqs (\ref{2fit}-\ref{4fit}) cannot be used to build three independent likelihoods. As we show in the companion paper~\cite{Quartin:2013moa}, a covariance matrix that takes into account the correlations can be computed. The covariance matrix contains also the errors for the various moments, which were computed without assuming a specific distribution and so make this approach robust against outliers.
Thanks to the use of the covariance matrix, a consistent likelihood approach is possible.

\subsection{On the lensing variance}

We turn our attention now to our estimate of the lensing variance, Eq.~\eqref{2fit}. We find quite smaller values as compared to the cosmology-independent Holz \& Linder fit of $\sigma_{HL}= 0.093 \, z$~\cite{Holz:2004xx}, which although outdated is still used for example by the \href{http://supernova.lbl.gov/Union/}{Supernova Cosmology Project} (responsible for the Union Compilations -- see~\cite{Amanullah:2010vv}). This is explicitly shown in the second row of Figure~\ref{fig:fits-sigma}. Our results, besides being in very good agreement with T11, are also in better agreement with the observational constraints obtained with real supernovae~\cite{Jonsson:2010wx} (although we point out that the error bars render their results statistically compatible with~\cite{Holz:2004xx}). This has important consequences that we detail below.

\begin{figure}[t!]
    \begin{centering}
    \includegraphics[width=.95\columnwidth]{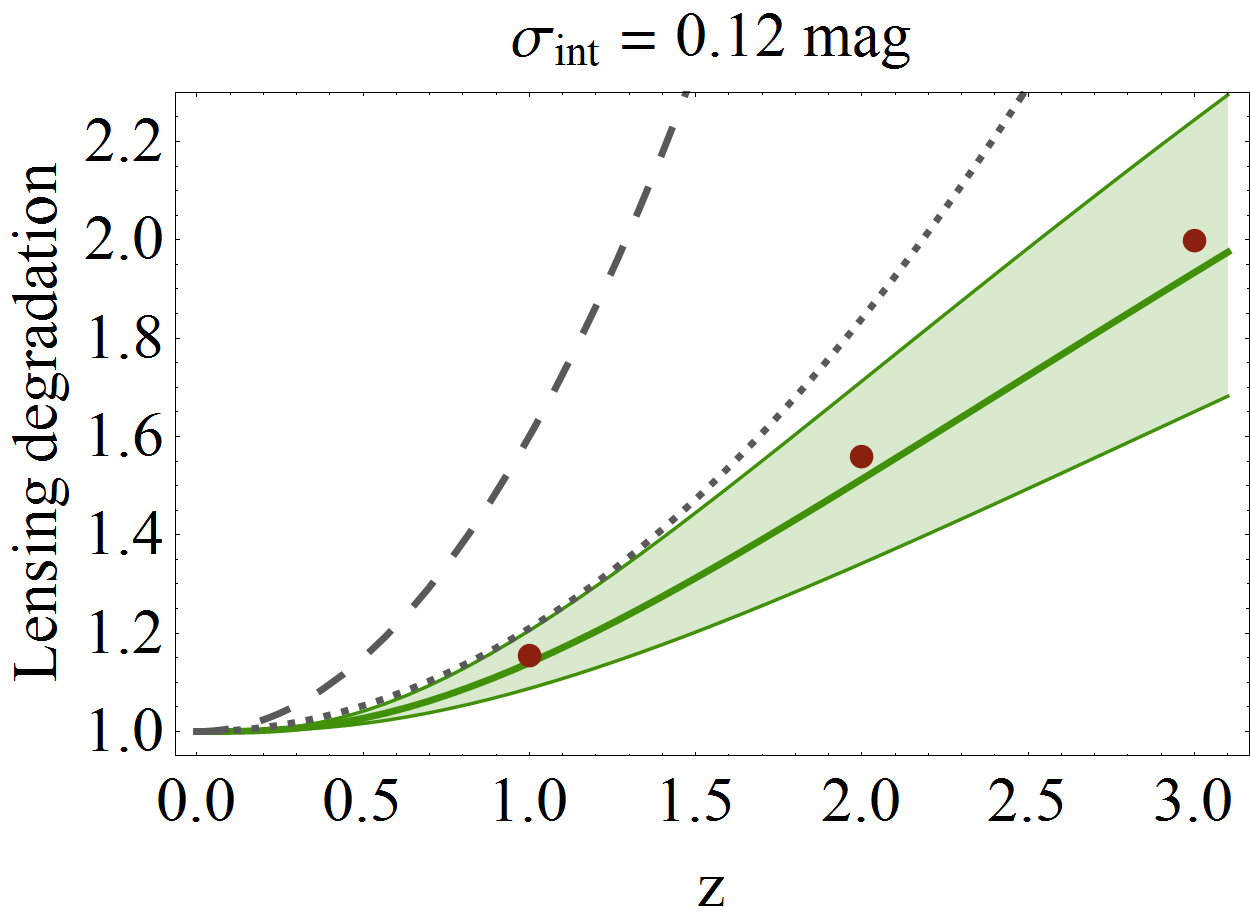}
    \caption{Standard candle lensing degradation factor as a function of redshift. \emph{Solid green:} our estimate for the best-fit parameters of WMAP9, together with the allowed region by changing the parameters by $2\sigma$. \emph{Dashed gray:} previous estimate found in~\cite{Holz:2004xx}. \emph{Dotted gray:} previous estimate found in~\cite{Jonsson:2010wx}. \emph{Big brown dots:} numerical N-body results found in T11~\cite{Takahashi:2011qd}. An intrinsic scatter of $\sigma_{\rm int} = 0.12$ mag is assumed.}
    \label{fig:lens-degradation} %
    \par\end{centering}
\end{figure}

In~\cite{Holz:2004xx} they estimate that due to the extra scatter of lensing supernovae quickly loses ``cosmological constraining power'' at high redshifts. For instance, assuming an intrinsic dispersion $\sigma_{\rm int} = 0.10$ mag, 3 supernovae would be needed at $z=1.5$ to provide the equivalent information of 1 supernova if there was no lensing. Here, we find that this ``degradation factor'', defined as
\begin{equation}
    1 + \frac{\sigma_{\rm lens}^2}{\sigma_{\rm int}^2}\,,
\end{equation}
is much smaller. Assuming $\Omega_{m0} = 0.3$ and $\sigma_8 = 0.8$, only 1.5 supernovae are needed to get the same information than what would otherwise be the case without lensing. If supernovae have larger intrinsic dispersion, say $\sigma_{\rm int} = 0.15$ mag, than the degradation is as low as 1.2. Figure~\ref{fig:lens-degradation} depicts this for the intermediate value of $\sigma_{\rm int} = 0.12$ mag, for the parameter values given by WMAP9~\cite{Hinshaw:2012aka}, together with the allowed region by varying the parameters by $2\sigma$.

We stress that this is very significant, as supernova observational strategies and survey-planning may be over-neglecting the power of high-$z$ candidates. Moreover, this alleviates one of the major issues for high-$z$ standard candle candidates, such as radio galaxies~\cite{Daly:2003iy}, GRBs~\cite{Ghirlanda:2004me,Basilakos:2008tp}, AGNs~\cite{Watson:2011um} or HII galaxies~\cite{Plionis:2011jj}, all of which can be observed at $z>3$. Since our fits fully take into account the dependence of lensing on cosmology, it allows a more careful estimate of the possible constraints of high-$z$ standard candles. We encourage, therefore, the scientific community to use our new fits for their own analyses, especially the fit of the lensing variance.

\section{Reconstructing the PDF from the moments}
\label{sec:logn}

Here we will show how to reconstruct the lensing PDF from the moments provided
by Eqs.~(\ref{2fit}-\ref{4fit}). We will make use of the known
fact~\cite{Taruya:2002vy} that the convergence distribution is well described by
the lognormal PDF.
This stems from the fact that the one-point distribution function of the matter
density field is itself well described by the lognormal
distribution~\cite{Coles:1991if, Kayo:2001gu} and that the convergence $\kappa$
is directly connected to the matter field $\delta_{M}$ as shown by
Eq.~(\ref{eq:kappa}).
The approach we will now take differs
from~\cite{Taruya:2002vy,Das:2005yb,Hilbert:2011xq}. We will now explain it in
detail.

\subsection{Lognormal distribution}

The lognormal distribution depends on the value of two parameters, $\mu_{\rm
gau}$ and $\sigma_{\rm gau}$, which are the mean and the dispersion,
respectively, of the gaussian distribution from which the lognormal is derived.
The lognormal PDF, together with its mean and variance, can be written as:
\begin{align}
f_{\rm logn}(x) &= \frac{\exp \left [{-\frac{(\log x - \mu_{\rm gau})^{2}}{2 \, \sigma_{\rm gau}^{2}}} \right]}{\sqrt{2 \pi} \, \sigma_{\rm gau} \, x} \,, \label{logn1} \\
\mu_{\rm logn} &= e^{ \mu_{\rm gau} + \sigma_{\rm gau}^{2}/2 } \,, \label{mulogn} \\
\sigma^{2}_{\rm logn} &= e^{2 \mu_{\rm gau} + \sigma_{\rm gau}^{2}} \left( e^{ \sigma_{\rm gau}^{2}} -1  \right) \,, \label{silogn}
\end{align}
where $x, \sigma_{\rm gau} >0$. One can invert Eqs.
\eqref{mulogn}--\eqref{silogn} so as to express $\mu_{\rm gau},\sigma_{\rm gau}$
as a function of the actual mean and dispersion of the lognormal distribution:
\begin{align}
\mu_{\rm gau} &= \frac{1}{2} \log \left[ \frac{\mu_{\rm logn}^{2}}{1 + \sigma_{\rm logn}^{2} / \mu_{\rm logn}^{2}} \right ] \,, \label{mugau} \\
\sigma_{\rm gau}^{2} &= \log \left( 1 + \sigma_{\rm logn}^{2} / \mu_{\rm logn}^{2} \right) \,. \label{sigau}
\end{align}
The lognormal distribution of Eq.~(\ref{logn1}) has a nonzero positive mean,
while the lensing distribution has an almost negligible mean $\bar \kappa$. So
as to model the convergence PDF we will then translate Eq.~(\ref{logn1}) into
our final template PDF for the convergence distribution:
\begin{align} \label{conpdf}
f_{\rm \kappa}(x) = \frac{\exp \left [{-\frac{[\log (x + \mu_{\rm logn}- \bar \kappa) - \mu_{\rm gau}]^{2}}{2 \, \sigma_{\rm gau}^{2}}} \right]}{\sqrt{2 \pi} \, \sigma_{\rm gau} \, (x + \mu_{\rm logn}- \bar \kappa)} \,,
\end{align}
where $\mu_{\rm gau},\sigma_{\rm gau}$ are given by Eqs.
\eqref{mugau}--\eqref{sigau} as a function of $\mu_{\rm logn},\sigma_{\rm
logn}$.

\subsection{Using information from the moments}
\label{sec:paramo}

\begin{figure*}
\begin{center}
\includegraphics[height=5.4 cm]{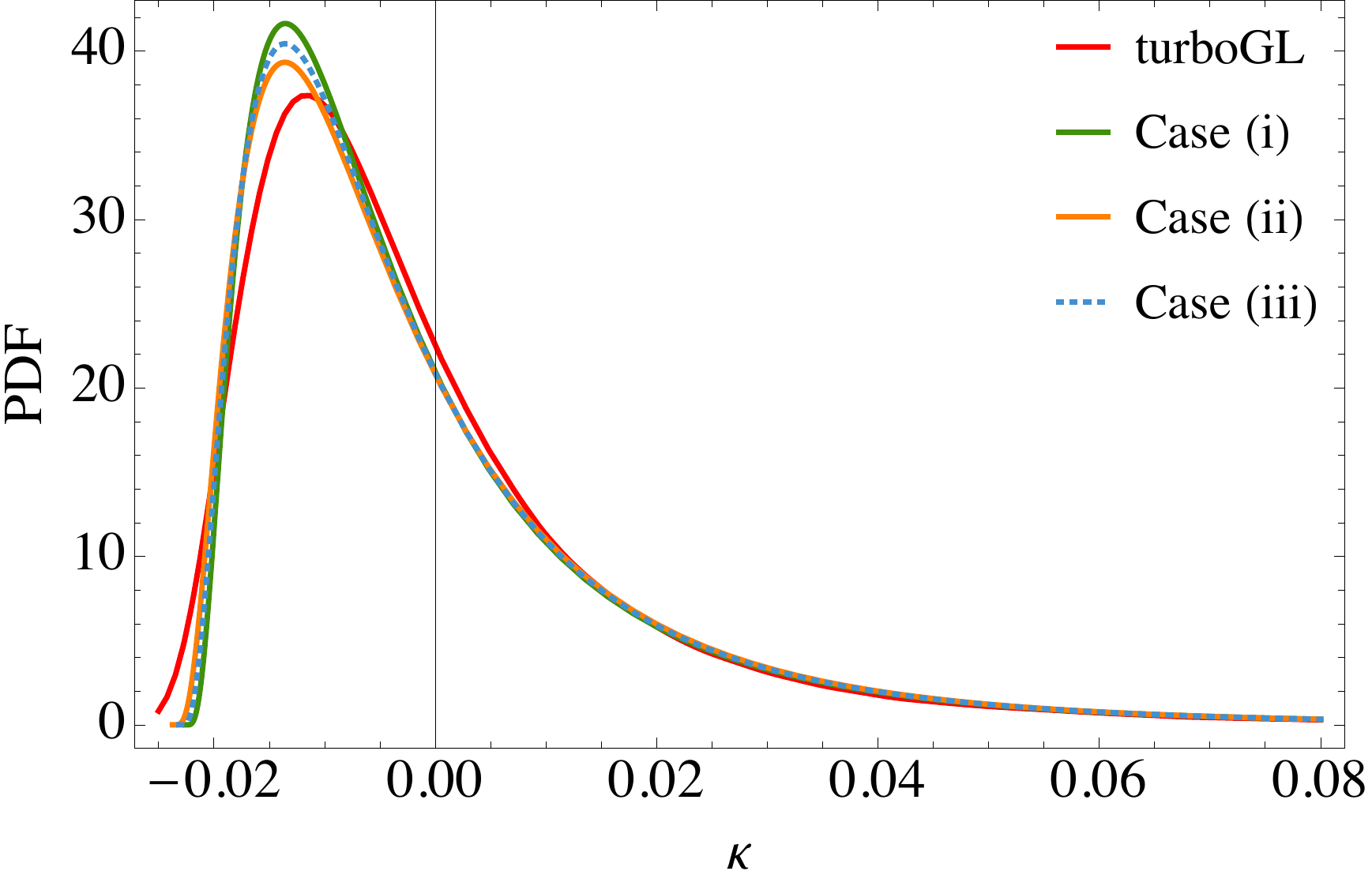}
\qquad
\includegraphics[height=5.4 cm]{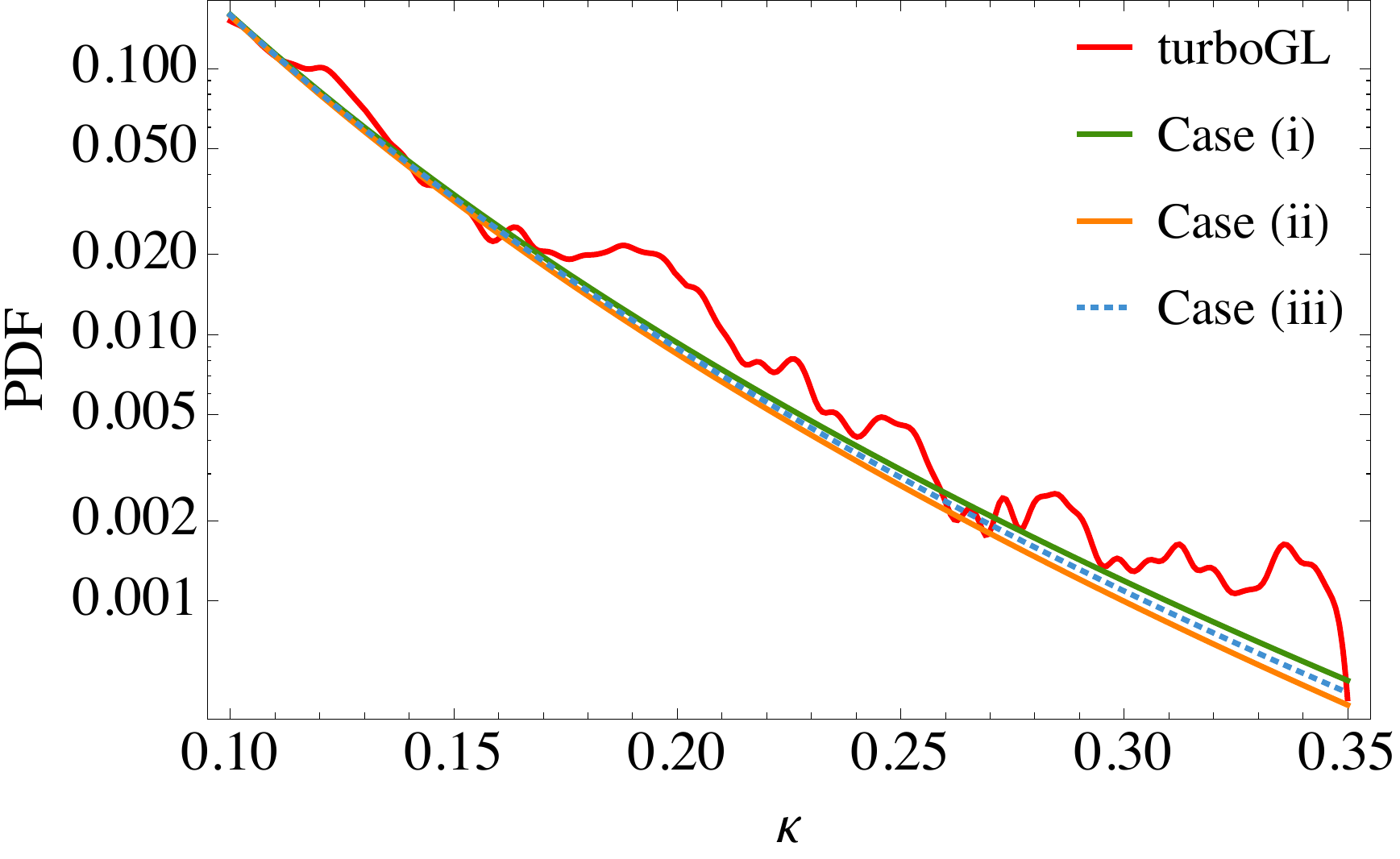}
\caption{Comparison of the reconstructed PDF's against the \tgl output.
\emph{Left}: the overall agreement is good as far as the body of the PDF is concerned.
\emph{Right}: the zoom on the high-magnification tail also shows a good level of agreement.
The numerical values of the parameters specifying the convergence PDF $f_{\rm \kappa}$ of Eq.~(\ref{conpdf}) are given in Table~\ref{logntab}. See Section \ref{sec:logn} for more details.}
\label{fig:logn}
\end{center}
\end{figure*}

The convergence PDF given in Eq.~(\ref{conpdf}) depends on three parameters, the
mean $\bar \kappa$, the dispersion $\sigma_{\rm logn}$ and the shape parameter
$\mu_{\rm logn}$ (as the mean has been set, $\mu_{\rm logn}$ effectively
constrains the higher moments of the $f_{\rm \kappa}$).
The dispersion $\sigma_{\rm logn}$ can be simply fixed using Eq.~(\ref{2fit}):
\begin{align} \label{lomapa1}
\sigma_{\rm logn} =  \frac{\log 10}{5} \; \sigma_{\rm lens} \,,
\end{align}
where the factor converts $\sigma_{\rm lens}$ from magnitudes back to convergences.
The mean can either be set to
\begin{align}
\bar \kappa =0 \,,
\end{align}
or to the (small) value given by
\begin{align}
\bar \kappa = -2 \, \sigma_{\rm lens}^{2}  \,,
\end{align}
so as to take in consideration some higher order corrections, see
Eq.~(\ref{eq:meank-vark-relation-2}) and related discussion.

We are then left with the shape parameter $\mu_{\rm logn}$, which can be used
either (i) to fix the third moment of $f_{\rm \kappa}$ to the $\mu_{3,{\rm
lens}}$ of Eq.~(\ref{3fit}) or (ii) to fix the fourth moment of $f_{\rm \kappa}$
to the $\mu_{4,{\rm lens}}$ of Eq.~(\ref{4fit}).
We need then to express the shape parameter $\mu_{\rm logn}$ as a function of
the third and fourth moments of $f_{\rm \kappa}$. This can be done for case (i)
and case (ii), respectively, by solving the following equations with respect to
$\mu_{\rm logn}$:
\begin{align}
\frac{\mu_{3,{\rm lens}}}{\mu_{\rm logn}^{3}} &= r^{4} \left (3 + r^{2} \right)  \,,  \label{shapeq3} \\
\frac{\mu_{4,{\rm lens}}}{\mu_{\rm logn}^{4}} &=  r^{4} \left (3 +16 r^{2} + 15 r^{4} + 6 r^{6} + r^{8} \right) \,,\label{shapeq4}
\end{align}
where $r=\sigma_{\rm logn}/\mu_{\rm logn} $.
The previous equations can be easily solved analytically. 
The algebraic expression of the solutions, though straightforward to obtain with
e.g.~Mathematica or Maple, are however rather long and not particular
illuminating. Consequently, they will not be explicitly reported here.
Alternatively, we will give ready-to-use fitting functions. The solutions to Eqs.
\eqref{shapeq3}--\eqref{shapeq4} depend on $\sigma_{\rm lens}$ and
$\mu_{3-4,{\rm lens}}$, which in turn depend (mainly) on
$\{z,\,\sigma_{8},\,\Omega_{m0}\}$. It, therefore, natural to use directly the
latter parameters. Fits to the solutions of \eqref{shapeq3}--\eqref{shapeq4}
with RMS errors better than 4\% within the ranges of \eqref{parange} are,
respectively:
\begin{widetext}
\begin{align}
\log_{10} \mu_{\rm logn}^{(3)} &= a (1.698\, -0.1979 c)-\frac{0.4081 b}{a-1.037}+\frac{0.2202}{a-1.037}-0.4081 b^3 c-0.4081 b^2-0.8128 b c+c-0.8128 \,, \label{shapeq3f} \\
\log_{10} \mu_{\rm logn}^{(4)} &=
-\frac{b-0.2861}{2.177 a+c-1.684}+1.721 a-0.4803 b^2-b c+1.156 c-0.7733 \,, \label{shapeq4f}
\end{align}
\end{widetext}
where $a= \log_{10} z$, $b= \log_{10} \sigma_{8}$ and $c= \log_{10} \Omega_{m0}$.

\subsection{Example}
\label{sec:example}

As an example of the formalism developed in the previous sections we will
consider the lensing PDF relative to the fiducial cosmology given by the 9-year
WMAP-only flat-$\Lambda$CDM best-fit values~\cite{Hinshaw:2012aka} and $z=1$.
In Figure~\ref{fig:logn} we compare the reconstructed PDFs against the \tgl
output. We considered three cases: (i) using for the shape parameter the value
given by (\ref{shapeq3}) (or equivalently (\ref{shapeq3f})) so as to match the
third moment, (ii) using the value given by (\ref{shapeq4}) (or equivalently
(\ref{shapeq4f})) so as to match the fourth moment and (iii) using the
intermediate value given by the mean of the last two. This makes sense as the
two values of $\mu_{\rm logn}$ are numerically close to each other.
To quantify the performance of the reconstructed PDF as far as the moments are
concerned, we list in Table~\ref{logntab} the error in the moments as compared
to the \tgl output. It is clear that $f_{\rm \kappa}$ is performing well,
particularly so in case (iii) where the errors are below the 2\% level, well within the modeling errors of $\sim$10\% intrinsic in our approach (see Appendix \ref{sec:errors}).
Also, even though the algebraic implementation of the lognormal distribution
presented here differs from e.g.~\cite{Hilbert:2011xq}, the overall shape of the
obtained distribution agrees rather well.

\begin{table}[h]
\begin{ruledtabular}
\begin{tabular}{cccccc}
Case & $\bar \kappa$ &$ \sigma_{\rm logn}$ & $\mu_{\rm logn}$         &   Error in $\mu_{3,{\rm lens}}$ & Error in $\mu_{4,{\rm lens}}$                     \\
\hline
(i) &0 & 0.0208 & 0.0225 & 0\% & 3.8\%       \\
(ii)& 0& 0.0208 &0.0237 & -2.3\% & 0\%       \\
(iii)&0& 0.0208 &  0.0231 & -1.2\% & 1.8\%
\end{tabular}
\end{ruledtabular}
\caption{Numerical value of the parameters specifying the convergence
distributions plotted in Figure~\ref{fig:logn}. The value of the shape parameter
$\mu_{\rm logn}$ for case (iii) has been obtained by averaging the values
relative to case (i) and (ii), which were obtained by matching the third and
fourth moments of $f_{\rm \kappa}$ to the lensing moments given by \tgl\!\!.}
\label{logntab}
\end{table}

\section{Conclusions}
\label{sec:conclusions}

Next generation supernova searches will collect thousands of SNIa lightcurves up to high redshift, vastly extending the present catalogs. In order to fully exploit this data abundance, the so-far almost negligible
error induced by lensing will need to be taken into account to avoid biasing the results. On the other hand, the lensing effect itself will become
a source of cosmological and astrophysical information, being related to the distribution and density profile of line-of-sight halos.

The crucial step in promoting lensing scatter from noise to signal is to
predict reliably its statistical properties, quantified through their magnification PDF. The calibration of the lensing PDF through $N$-body simulations is accurate but extremely time consuming and cannot, at this time, be carried out for a sufficiently large set of cosmological models as the one needed in the reconstruction of the SNIa likelihood.

In this paper we use the fast computation allowed by the sGL method~\cite{Kainulainen:2009dw,Kainulainen:2010at,Kainulainen:2011zx} -- which distributes halo and sources and evaluates the resulting lensing effect in the linear regime in a matter of seconds, i.e.~several orders of magnitude faster than any $N$-body approach -- to reconstruct the lensing PDF for an array of cosmological models.
First, we have shown that our fast results match those from state-of-the-art $N$-body simulations.
Then, we illustrated that the PDF is practically insensitive to $h,\omegak, w, n_{s}, \Omega_{b0}$, while strongly dependent on $\sigma_8,\Omega_{m0}$, in agreement with expectations.
We have quantified the latter dependence by plotting the behavior of the second, third and fourth central moments.
We then provided accurate fits that give the moments for any redshift and any $\sigma_8,\Omega_{m0}$ within a reasonable range. These fits have the advantage of being directly employable in a parameter estimation likelihood analysis. We quantify the error of the fits with respect to the $N$-body simulations to be at the $\sim$10\% level. We also find that a shifted log-normal PDF -- of which we give an easy-to-use prescription -- approximates relatively well the numerical PDF.

Our analysis considerably improves upon the popular fit~\cite{Holz:2004xx}. In particular, we find  that our estimate of the additional variance induced by lensing is quite smaller than in Ref.~\cite{Holz:2004xx}. The lensing noise is found to be comparable to the intrinsic scatter at redshift $z\approx 3$ rather than at $z\approx 1.3$ as in Ref.~\cite{Holz:2004xx}, for a WMAP9 cosmology. This not only is in better agreement with preliminary observational constraints of the lensing PDF~\cite{Jonsson:2010wx} but also makes high-redshift supernova observations more relevant for cosmology and may entice reviews of supernova observational strategies.

It is clear that the next step in this line of research is to invert the problem and ask how well one can use the SNIa PDF to estimate the cosmological parameters, in particular $\sigma_8,\Omega_{m0}$. If one assumes the SNe magnitudes to be intrinsically Gaussian, then the entire observed third order moments and the excess-over-Gaussian fourth order moment are a result of lensing. If the rare strongly non-linear lensing events can be identified and eliminated from the analysis then the observations can be directly compared to our results and employed to constrain $\sigma_8,\Omega_{m0}$, directly from lensing. On the other hand, if we fix the cosmological model and the observations show a significant deviation from the predicted lensing effect, then one would have an evidence for an intrinsic non-Gaussianity of the sources or a  violation of lensing linearity. This work will be carried out in a subsequent paper.

\begin{acknowledgments}
It is a pleasure to thank Stephan Hilbert, Kimmo Kainulainen, Eric Linder, Martin Makler, Bruno Moraes, Ribamar Reis, Ignacy Sawicki, Brian Schmidt, Peter Schneider and Ryuichi Takahashi for fruitful discussions. LA and VM acknowledge support from DFG through the TRR33 program ``The Dark Universe''. MQ is grateful to Brazilian research agencies CNPq and FAPERJ for support and to ITP, Universität Heidelberg for hospitality during part of the development of this project.
\end{acknowledgments}

\appendix

\section{Properties of the weak-lensing PDF}

\label{sec:app-review}

In this work we are implicitly making use of the multiple-lens-plane
approximation, which has been proven to be an exquisitely good one
in all cases of physical interest for cosmology. In such approximation,
it is imperative to establish which is the plane of reference: the
source plane or the image plane. In fact, the statistical properties
of the distributions of weak-lensing distortions vary significantly
depending on this choice, as we will discuss below. \emph{In a broad sense, the source plane is closely related to observable quantities, whereas the image plane is often more directly related to numerical simulations.} We shall denote by a subindex {}``S'' ({}``I'') averages conducted on the source (image) plane.

A classic and general result of gravitational lensing is that photon
conservation implies that the magnification~($\bar{\mu}$) PDF has unitary mean. This is true even outside the weak lensing regime,
but this result assumes implicitly that one is working on the source
plane. In other words, $\,\langle\bar{\mu}\rangle_{{\rm S}}\,\equiv\,1\,\neq\,\langle\bar{\mu}\rangle_{{\rm I}}$.

Since magnification $\bar{\mu}$, convergence $\kappa$ and shear
$\gamma$ are related by
\begin{equation}
\bar{\mu}=\frac{1}{(1-\kappa)^{2}-\gamma^{2}}\,,
\end{equation}
 one has for small $\kappa$
 \begin{equation}
\bar{\mu}\simeq1+2\kappa+3\kappa^{2}+\gamma^{2}\,.\label{eq:mu-weak}
\end{equation}
 This implies that
 \begin{align}
 & \langle\bar{\mu}\rangle_{{\rm S}}\equiv1\simeq1+2\langle\kappa\rangle_{{\rm S}}+3\langle\kappa^{2}\rangle_{{\rm S}}+\langle\gamma^{2}\rangle_{{\rm S}}\label{eq:mean-mu-weak}\\
 & \therefore\quad\langle\kappa\rangle_{{\rm S}}\simeq-\frac{3}{2}\langle\kappa^{2}\rangle_{{\rm S}}-\frac{1}{2}\langle\gamma^{2}\rangle_{{\rm S}}\,.\label{eq:meank-vark-relation}\end{align}
Now, in the weak-lensing approximation one has $\langle\kappa^{2}\rangle_{{\rm S}}=\langle\gamma^{2}\rangle_{{\rm S}}$~\cite{Takahashi:2011qd}.
Since \tgl does not compute the shear, we made use of the PDF's
derived in~\cite{Takahashi:2011qd} to test this numerically and
found that, at $z=1$, $\langle\kappa^{2}\rangle_{{\rm S}}=4.7\;10^{-4}$
and $\langle\gamma^{2}\rangle_{{\rm S}}=5.3\;10^{-4}$, an agreement
to the 10\% level. Using this equality, the above simplifies to
\begin{align}
\langle\kappa\rangle_{{\rm S}}\,\simeq\,-2\langle\kappa^{2}\rangle_{{\rm S}}\,.\label{eq:meank-vark-relation-2}
\end{align}
 This correlation between mean and variance was indeed observed explicitly
in \cite{Takahashi:2011qd} (see their Figure 5).

If one repeats the above calculations in the image plane, results
differ significantly. In the image plane there is a bias towards higher
magnification events. This implies that \begin{align}
\langle\kappa\rangle_{{\rm I}} & =\langle\kappa\mu\rangle_{{\rm S}}\simeq\langle\kappa\rangle_{{\rm S}}+2\langle\kappa^{2}\rangle_{{\rm S}}+3\langle\kappa^{3}\rangle_{{\rm S}}+\langle\kappa\gamma^{2}\rangle_{{\rm S}} \,.
 \end{align}
Using \eqref{eq:meank-vark-relation-2} and up to higher
order corrections, one finds that \begin{align}
\big|\langle\kappa\rangle_{{\rm I}}\big|\,\sim\,{\cal O}\big(\langle\kappa^{3}\rangle_{{\rm S}}\big)\,\ll\,\big|\langle\kappa\rangle_{{\rm S}}\big|\,.\end{align}
 Using once again the PDFs derived in \cite{Takahashi:2011qd} we
find that, at $z=1$, $\langle\kappa\rangle_{{\rm I}}\simeq-0.005\langle\kappa^{2}\rangle_{{\rm I}}$,
which should be confronted with~\eqref{eq:meank-vark-relation-2}.

\begin{figure}[t!]
    \begin{centering}
    \vspace{.0cm}
    \includegraphics[width=\columnwidth]{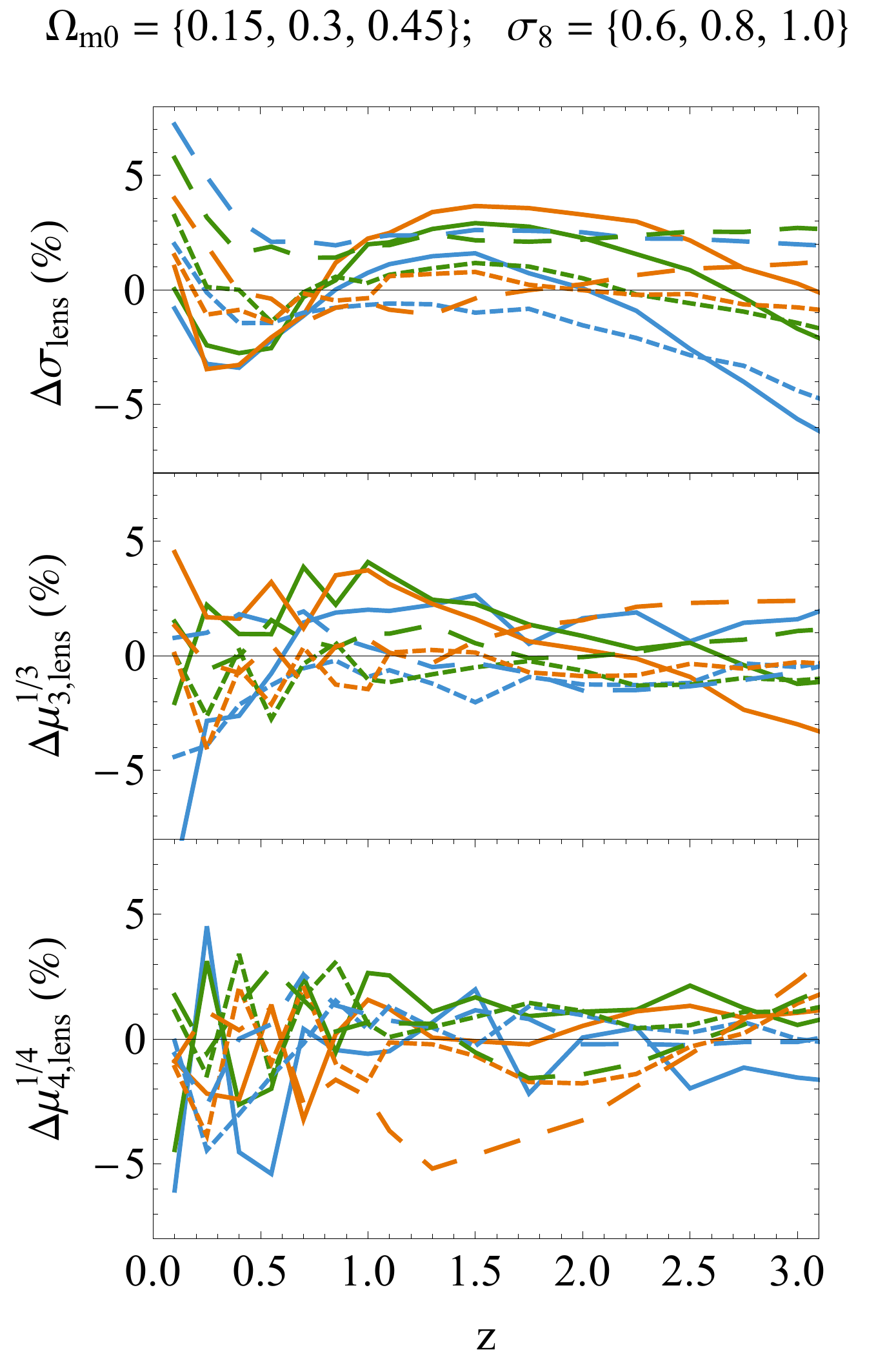}
    \caption{Percentage differences between fits   and numerical results, showing that the fit is almost always within 5\% of the numerical values, in agreement with the quoted average RMS error of 4\% in the whole domain of Eq.~\eqref{parange}. \{Full,  Dashed, Long-dashed\} lines stand for $\Omega_{m0} = \{0.15,\,0.3,\,0.45\}$, while \{Blue, Green, Orange\} lines stand for $\sigma_8 = \{0.6,\,0.8,\,1.0\}$.}
    \label{fig:fits-errors} %
    \par\end{centering}
\end{figure}

Another important aspect of the weak-lensing PDF is the theoretical
prediction of how the PDF of $\bar{\mu}$ behaves for $\bar{\mu}\gg1$.
Catastrophe theory predicts that, for a point source, in the high
magnification limit one has~\cite{Hilbert:2007ny,petters2001singularity}
\begin{align}
{\rm PDF}(\bar{\mu})_{{\rm S}}\;\propto\;\mu^{-3}\,,\\
{\rm PDF}(\bar{\mu})_{{\rm I}}\;\propto\;\mu^{-2}\,.\end{align}
 This in turn implies that even in the source plane, for a point source
all central moments above 1 would be infinite. In other words, the extra
variance in a standard candle due to lensing would be unbounded. For
an extended source this is no longer the case and the PDF is suppressed
for $\bar{\mu}>\bar{\mu}_{{\rm max}}$. Now $\bar{\mu}_{{\rm max}}$
is typically well beyond the weak-lensing validity range, even for
galactic-scale sources. It was shown in~\cite{Takahashi:2011qd}
that for a circular source of radius $10$~kpc/h at $z=5$, $\bar{\mu}_{{\rm max}}\simeq4$;
for a supernova $\bar{\mu}_{{\rm max}}$ must be much higher. Thus
if one wants to make use of the weak-lensing central moment predictions,
one must enforce a lower cut $\bar{\mu}_{{\rm max}}^{{\rm lower}}$
to the PDF, which can in practice be achieved by removing all sources
for which there is strong evidence that $\,\bar{\mu}>\bar{\mu}_{{\rm max}}^{{\rm lower}}$ (see Section~\ref{compase}).


\section{Error budget of our results}
\label{sec:errors}

\begin{table}[t!]
\begin{ruledtabular}
\begin{tabular}{lr}
Type         &   Error in $\mu_n^{1/n}$                      \\
\hline
Weak-lensing approximation & 5\%        \\
Halo model & 5\%        \\
Mass function          & 3\%                       \\
Concentration parameters   & 3\% \\
Our fitting functions  & 4\% \\
\hline
Total error (without fitting error) & $\simeq$8\% \\
Grand total error & $\simeq$10\%
\end{tabular}
\end{ruledtabular}
\caption{Summary of the errors for $\mu_n^{1/n}$ intrinsic in the approach of this paper, for $n=\{2,\,3\}$. For $n=4$, the first four errors are larger, giving a total of the order of $10-15\%$}
\label{errors}
\end{table}

Our results are subjected to modeling errors. We we will quantify them as far as $\sigma_{\rm lens} (= \mu_{2,{\rm lens}}^{1/2})$, $\mu_{3,{\rm lens}}^{1/3}$ and $\mu_{4,{\rm lens}}^{1/4}$ are concerned.
The first source of error comes from the weak-lensing approximation, which we estimated through comparison with exact results to be of about 5\%. The second source is the matter model adopted. The halo model neglects indeed extended structures such as filaments and walls. Based on the results of \cite{Kainulainen:2010at}, we estimated this error at the 5\% level. There are then the errors inherent to the mass function~\cite{Jenkins:2000bv} and concentration parameter model~\cite{Zhao:2008wd} used. By changing the latter within the error ranges stated in the respective papers we found that they add a 3\% error each to the overall error budget. Finally, there is the 4\% error due to the fitting functions. Figure~\ref{fig:fits-errors} gives the percentage differences between fits and numerical results, i.e.
\begin{equation}
    \Delta\mu_n^{1/n}(\%)\,\equiv\,\frac{\mu_n^{1/n,\,{\rm fit}}-\mu_n^{1/n,\,{\rm numerical}}}{\mu_n^{1/n,\,{\rm numerical}}}
\end{equation}
showing that for all three moments the fits are almost always within 5\% of the numerical values.

We list in Table~\ref{errors} a summary of the errors intrinsic in the present modeling, which together do not exceed the 10\% level. These figures mainly refer to the second and third moment; the fourth moment is subjected to a larger error of the order of $10-15\%$. Finally, though the fitting functions are accurate for the full ranges given in (\ref{parange}), the first four errors listed in Table~\ref{errors} are only accurate for lensing results in a slightly narrower range: $0\le z \le 2$, $0.6 \le \sigma_{8} \le 1.0$ and $0.1 \le \Omega_{m0} \le 0.5$.

\bibliography{cosmo-lensing}


\end{document}